\documentclass[prl,twocolumn,showpacs]{revtex4}
\usepackage{epsfig}
\usepackage{amssymb,amsmath,tabularx,trees}
\usepackage[latin1]{inputenc}
\usepackage{wrapfig}
\usepackage{subfigure}
\usepackage{bm}

\begin{document}
\title{Stress-driven phase transformation and the roughening of solid-solid interfaces}

\author{L.~Angheluta$^1$, E.~Jettestuen $^1$,
J.~Mathiesen$^1$, F.~Renard$^{1,2}$, B.~Jamtveit$^1$}
\affiliation{$^1$ Physics of Geological Processes, University of Oslo, Oslo, Norway  \\ $^2$ LGCA-CNRS-OSUG, University of Grenoble, BP 53, F-38041, France}
%\email{luizaa@fys.uio.no}
\date{\today}
% ------------------------------------------------------------------------ %%
%
%%%% ABSTRACT
%
% ------------------------------------------------------------------------ %%

\begin{abstract}
The application of stress to multiphase solid-liquid systems often results in morphological instabilities. Here we propose a solid-solid phase transformation model for roughening instability in the interface between two porous materials with different porosities under normal compression stresses. This instability is triggered by a finite jump in the free energy density across the interface, and it leads to the formation of finger-like structures aligned with the principal direction of compaction. The model is proposed as an explanation for the roughening of stylolites - irregular interfaces associated with the compaction of sedimentary rocks that fluctuate about a plane perpendicular to the principal direction of compaction.
\end{abstract}

%%%% ---------------------------------------------------------------------- %%
%
%  BEGIN LETTER
%
%%%  ---------------------------------------------------------------------- %%

\maketitle

%%%% ---------------------------------------------------------------------- %%
%
%  TEXT
%
%%% ------------------------------------------------------------------------ %%

%\section{Introduction}
Morphological instabilities in systems out of equilibrium are central to most research on pattern formation. 
A host of processes give rise to such instabilities, and among the most intensively studied are the surface 
diffusion mediated Asaro-Tiller-Grinfeld instability~\cite[]{Asaro72}, \cite[]{Grinfeld86}, 
\cite[]{Srolovitz88} in the surfaces of stressed solids in contact with their melts, 
surface diffusion mediated thermal grooving and solidification controlled by thermal diffusion in the bulk melt
\cite[]{Mullins57}. In sedimentary rocks and other porous materials local stress variations typically promote 
morphological changes via dissolution in regions of high stress, transport through the fluid saturated pore space 
and precipitation in regions of low stress. This phenomenon is known as pressure solution or chemical compaction.
Such processes are often accompanied by the nucleation and growth of thin irregular sheets, interfaces or seams called stylolites
~\cite[]{Weyl59}. Stylolites form under a wide range of geological conditions as rough interfaces that
fluctuate about a plane perpendicular to the axis of compression. They are common in a variety of rock types, 
including limestones, dolomites, sandstones and marbles, and they appear on scales ranging from 
the mineral grain scale to meters or greater. A common feature of stylolitic surfaces is small scale roughness combined with large vertical steps in the 
direction of the compression. Residual unsoluble minerals (i.e. clays, oxides) often accumulate at the interface as stylolites evolve.
Despite the considerable attention given to the rich morphology of stylolites 
there is still no consensus on the mechanism(s) controlling their formation \cite[]{Oelkers96}, \cite[]{Schmittbuhl04}, \cite[]{Aharonov07}, \cite[]{Koehn07}.
Here we demonstrate that even if the stylolite is a consequence of pressure solution 
alone, porosity or other material property gradients may drive the roughening process. In particular, we demonstrate that a compressional 
load normal to a solid-solid phase boundary gives rise to a morphological instability.

Generally, rocks are heterogeneous bodies with spatially variable porosities. The strain energy densities may be
larger in regions of high porosity (i.e.~low modulus) than in regions of low porosity. Thermodynamically, the 
total free energy of the system can be reduced by reducing the porosity variations. In this letter, a simple 
model for stylolite formation, in which high porosity rock is transformed into low 
porosity rock at the interface between the low porosity and high porosity materials, is investigated. This solid-solid ``phase 
transformation'' is driven by the free energy density gradients, which can be substantial in regions with large porosity variations. The general approach used in this work could be applied to other solid-solid interface roughening phenomena.
%------------------ Figure 1 -----------------------------------
\begin{figure}
\centering
\includegraphics[width = 0.42\textwidth, angle = 0]{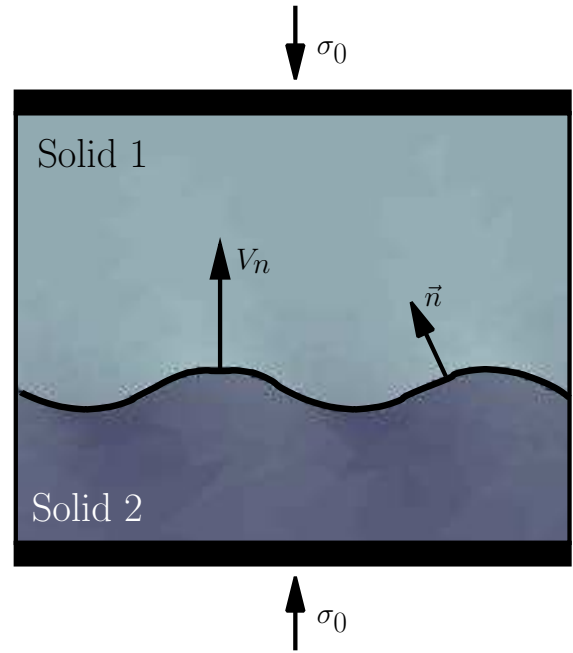}

\caption{(Color online) Basic setup of the model for a moving interface between two elastic solid phases, characterized by
different Young's moduli ($E_2>E_1$) and Poisson's ratios $\nu_{1}, \nu_{2}$. The interface boundary propagates with a normal velocity
$V_n$, when the solids are subjected to uniform far-field compressional stresses $\sigma_0$ in the vertical direction.}
\label{fig-setup}
\end{figure}

%\section{Solid-solid interface}\label{Sec.2}
We consider a two-dimensional system divided into elastic regions with different but homogeneous porosities (Fig.~\ref{fig-setup}).
Without lack of generality, we limit our consideration to two dissimilar materials separated by a single interface.
The stress boundary condition is a uniform compression
in the vertical direction applied at the top and bottom boundaries.
The two phases are separated by a sharp and coherent boundary, i.e. no defects or voids can form along the interface.
This translates into continuity of the displacement
vector $\textbf u(\textbf r,t)$ across the interface,
\begin{equation}
  [\textbf{u}] =0.
\end{equation}
Here, and in other equations the brackets denote the jump in the quantity inside the brackets when 
the interface is approached from above and below. Under given load conditions the displacement field induced 
by the compression gives rise, in the linear regime, to a strain tensor of the form
\begin{equation}
  \label{eq:linela}
 \epsilon_{ij} = \frac 1 2\left (\frac {\partial u_j}{\partial x_i}+\frac
 {\partial u_i}{\partial x_j}\right).
\end{equation}
The two solid phases are characterized by their Young's moduli, $E_{1,2}$,
and Poisson's ratios, $\nu_{1,2}$. When $E_1<E_2$, the upper region will be compressed the most 
and therefore the elastic energy density will be higher in this region. The first step towards a model for 
the roughening of a solid-solid interface is based on this simple observation. First the elastic 
parameters of the materials are related to their porosities. Luo and Weng~\cite{Luo87} proposed
a homogenization method relating the effective bulk and shear moduli to the porosity of the solid. 
Using this approach, the effective Young's modulus decreases monotonically with the porosity.
Consequently, a finite jump in porosity across the interface induces a jump in the elastic energy density, which drives
the motion of the interface. Thermodynamically, the evolution of the interface corresponds to a phase 
transformation from a high to a low energy state.

It is assumed that the phase transformation occurs on a time scale that is much longer that the time required for elastic waves to propagate across the system, and the system is therefore always in elastostatic equilibrium. For an isotropic and homogeneous
 elastic body the elastic equilibrium condition is given by
\begin{equation}
  \label{eq:equil}
  \frac{\partial \sigma_{ij}}{\partial x_j}=0,
\end{equation}
together with the uniform uniaxial compression stress boundary condition.
\begin{equation}
\sigma_{ij}(x,y=\infty) = \sigma_0\delta_{i,y}\delta_{j,y}<0,
\end{equation}
and the stress jump across the curved interface due to the effective surface tension is given by  
\begin{equation}
[\sigma_{ij} n_j] = -\gamma\kappa n_i\textrm{ at the interface }
\Gamma_t\label{eq.ForceBal2},
\end{equation}
where $\kappa$ is the curvature and $\gamma$ is the local surface
tension. In the limit of negligible surface tension, the
stress vector is continuous across the interface ($[\sigma_{ij} n_j] =0$).

For completeness, the basic principles used to derive an equation of motion are presented. When the system
approaches an equilibrium configuration, the free energy
will be a non-increasing function of time:
\begin{equation}\label{eq.GEnBal}
\frac{d}{dt}\bigg(\int_{V}\mathcal{F}dv+\int_{\Gamma_t}\tilde{\mathcal{F}}ds\bigg)\le
0,
\end{equation}
where $\mathcal{F}$ is the free energy per unit volume and $\tilde{\mathcal{F}}$ is
the interfacial free energy per unit area. Here, the subscript $V$ indicates a volume integration and
$\Gamma_t$ indicates integration over the interface. The interfacial energy dissipation is obtained by
confining the domain of integration to a narrow zone along the
interface and taking the zero thickness limit \cite[]{Gurtin91}. This gives 
\begin{equation}
-\int_{\Gamma_t}[\mathcal{F}]V_nds+
\int_{\Gamma_t}\bigg(\frac{d\tilde{\mathcal{F}}}{d
t}-\kappa\tilde{\mathcal{F}}V_n\biggr) ds\le 0,
\end{equation}
where $V_n$ is the normal velocity and $\kappa$ is the local
curvature of the interface. This implies the differential form
given by
\begin{equation}
\frac{d\tilde{\mathcal{F}}}{dt}
-(\kappa\tilde{\mathcal{F}}+[\mathcal{F}])V_n\le 0,
\end{equation}
where the first term is the total time derivative of the local
interfacial energy density. The local free energy is a function of
the surface tension only (like fluid-solid interfaces) and
thus, it is independent of time. Therefore, the time
derivative can be neglected leading to the inequality
\begin{equation}\label{eq:EnergyBal}
-(\kappa\tilde{\mathcal{F}}+[\mathcal{F}])V_n\le 0.
\end{equation}
In the linear response regime, this inequality is satisfied when the velocity (the thermodynamic flux) is a
linear function of the driving force,
$(\kappa\tilde{\mathcal{F}}+[\mathcal{F}])$, namely
\begin{equation}\label{dynlaw}
V_n\sim c\bigg(\kappa\tilde{\mathcal{F}}+[\mathcal{F}]\bigg),
\end{equation}
with $c\ge 0$. In the absence of surface tension, the normal
velocity is simply  proportional to the jump in the strain energy,
i.e. $V_n \sim [\mathcal{F}]$. While the dynamical law for the
interface Eq. (\ref{dynlaw}) is very simple, the implementation in a numerical model is more challenging.

%\section{Numerical model}\label{Sec.3}
The model (Fig.~\ref{fig-setup}) of the moving solid-solid interface was numerically implemented using the local 
force balance and energy dissipation equations (Eqs.~(\ref{eq.ForceBal2}) and (\ref{dynlaw})). The stress 
field is obtained using the Galerkin finite element discretization of the elastostatic equations and the phase boundary is captured using the level set method. The level set method (\cite{Sethian99},
\cite{Osher03}) is a powerful and reliable technique for tracking surfaces
in any number of dimensions. At any time $t$, the d-dimensional interface
$\Gamma_t$ may be defined as the zero level cut through a scalar field $\varphi$ (d+1-dimensional surface),
namely
\begin{equation}\label{eq:ls1}
  \varphi(\textbf{x},t) = 0\textrm{ , where }\textbf{x}\in \Gamma_t.
\end{equation}
A change in the zero-level cut in response to a change in the scalar field may then be interpreted as a motion
of the interface. Therefore the change in the scalar field must correspond to motion of the zero-level cut 
with a given normal velocity $\textbf{V}$, this is done by updating the scalar field using a simple advection-like 
equation
\begin{equation}\label{eq:ls3}
  \frac{\partial\varphi}{\partial t} + V_\text{n}|\nabla\varphi| = 0.
\end{equation}
The advection of the level set function is solved on a separate lattice using an
upwind-scheme. The full dynamical model of the solid-solid phase transformation front is then given by this 
equation together with Eq. (\ref{dynlaw}).

%------------------ Figure 2 -----------------------------------
\begin{figure}
\centering \epsfig{width = 0.45\textwidth, angle = 0, file=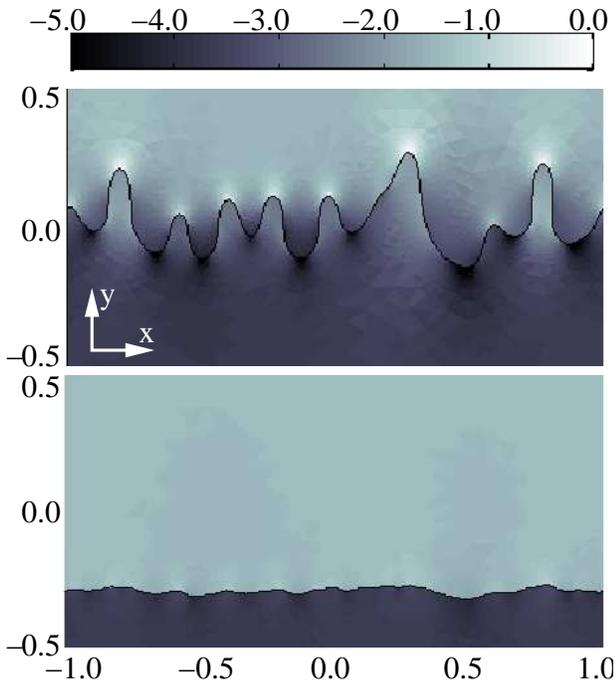}

\caption{(Color online) Map of the logarithm of the elastic energy in the solids
during the roughening process, with $E_{1}=10$ GPa, $E_{2}=60$ GPa, $\nu_{1}=\nu_{2}=0.3$, and $\sigma_{0}=0.05$ MPa.
Lower panel: initial $h(x)$ at $t=0$. Upper panel: interface at a later stage
of roughening.} \label{fig-2D-energy}
\end{figure}
%
%------------------ Figure 3 -----------------------------------
\begin{figure}
\centering
\includegraphics[width = 0.45\textwidth, angle = 0]{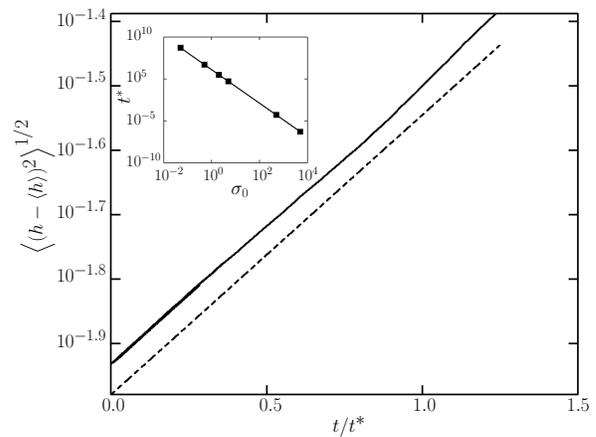}

\caption{Roughness as a function of time for six different external
compression stresses $\sigma_0$ for a fixed jump in the Young's modulus and zero surface tension.
The root means square height is plotted as a function of time, rescaled with the characteristic
roughening time $t^*$, on a semi-logarithmic scale. The data collapse shows the exponential
roughening of the interface $\exp(t/t^*)$ with a stress independent pre-exponential factor.
Inset: the characteristic time as a function of external stress is
given by $t^*\sim \sigma^{-2}_{0}$. } \label{fig-roughness-sigma}
\end{figure}

%\section{Discussion of results}\label{Sec.4}
The numerical simulations were started with a random interface generated by a directed random walk (Fig.~\ref{fig-2D-energy}, lower panel).
The temporal evolution of the interface (Fig.~\ref{fig-2D-energy}, upper panel) was then recorded for different external stresses, 
$\sigma_0$, and elastic constants ($E$, $\nu$). The elastic constants were computed from homogenization relations between elasticity and
porosity \cite[]{Luo87}.

Initially, the roughness of the interface grows exponentially
\begin{equation}
\sqrt{\int_{\Gamma_t}\bigg(h(s,t)-\overline{h}(t)\bigg)^2ds}\sim
\exp(t/t^*),
\end{equation}
with a characteristic roughening time $t^*$ that depends on the external stress and the jump in the elastic
properties. To estimate the functional dependence of $t^*$, a set of numerical simulations was performed. 
First, the external stress $\sigma_0$ was systematically varied between 0.05 and 50 MPa for fixed values of the elastic constants
($E_1=40$ GPa, $E_2=60$ GPa, $\nu_{1}=\nu_{2}=0.3$).

\begin{figure}
\centering
\includegraphics[width = 0.45\textwidth, angle = 0]{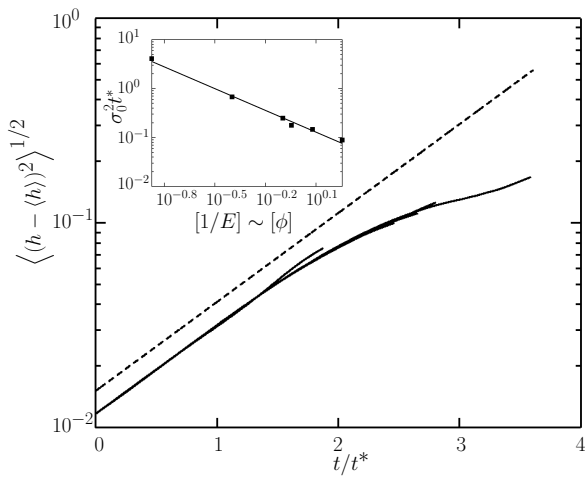}

\caption{Roughness as a function of time for different jumps in the Young's modulus or 
porosity $[1/E] = 1/E_1-1/E_2\sim[\phi]$ for six different external stresses and zero surface tension.
The data is plotted against the time rescaled with $t^*$ estimated numerically
from the initial exponential growth. At large times, the roughness
grows quadratically with increasing time. Inset: Log-log plot of the characteristic
time rescaled with the corresponding external stress
$\sigma_0^2\cdot t^*$ as a function of the jump $[1/E]$. $t^*\sim
\sigma_0^2\cdot [1/E]^{-\alpha}$, with the scaling exponent
$\alpha\sim 1.36$.} \label{fig-roughness-DE}
\end{figure}

The results shown in Fig.~\ref{fig-roughness-sigma} suggests that the characteristic time scales as 
$t^*\sim \sigma_0^{-2}$, and the prefactor depends on the elastic properties. In order to investigate this 
type of relation, the elastic constants across the interface were varied ($E_{2}=50$ GPa, $E_{1}$ in the
range 5-16 GPa, $\nu_{1}=\nu_{2}=0.3$). The data for the interface 
roughening collapses onto a curve which is exponential at small time-scales with a cross-over to a quadratic 
form at larger times (Fig.~\ref{fig-roughness-DE}). Extracting the characteristic time in the exponential 
growth regime and rescaling it with $\sigma_0^2$, the functional dependence with 
respect to the jump in the elastic constants across the interface was determined
(see Fig.~\ref{fig-roughness-DE}, inset). The cross-over from exponential to algebraic roughening depends on the value of the jump. For large jumps in the elastic energy or porosity, 
the roughening quickly undergoes a transition from exponential to quadratic growth in time. 
This cross-over time is related to the formation and growth of the
finger-like structures shown in Fig.~\ref{fig-2D-energy}. By a simple dimensional argument, we also determined that
the generic dependence of $t^*$ on the external stress and elastic constants, has the form
\begin{equation}
t^*\sim\frac{L}{V_n}\sim\frac{L}{c[\mathcal{F}]}\sim\frac{L}{c}
\frac{1}{[(1-\nu^2)/E]} \frac{1}{\sigma_0^2},
\end{equation}
where $[\mathcal{F}] = \mathcal{F}_1-\mathcal{F}_2$ is the jump in the free energy density across the
interface. The scaling relation is consistent with the numerical simulation results.
%
%------------------ Figure 4 -----------------------------------
The stress field may be calculated analytically with the same boundary and interface conditions for a straight 
interface perturbed by a sine function with small amplitude $A$. In linear perturbative analysis 
($A<\!\!< 1$) without surface tension, the solution is obtained using the method of Airy's potentials 
for 2D elastostatics \cite[]{Gao91}. The energy jump $[E]$ is proportional to $k A^2$, where $k =2\pi/\lambda$
is the wave number. In other words, all the modes are unstable and those with
the smaller wavelengths grow faster in the linear regime. The surface tension adds an ultraviolet cut-off
resulting in small scales smoothening of the interface. The system was tested with and without surface tension and 
in both cases the qualitative behavior was the same - an initial exponential roughening with a crossover to a
finger-formation regime. Eventually, these fingers may stabilize due to transport of dissolved minerals and precipitation leading to pore clogging. 
Stress concentration at the tips is an important characteristic of the system.
In stylolites the roughening is often accompanied by small fractures aligned with the direction of compression, 
and this may be explained by the model if the stress concentration at the fingers exceeds the yield strength of the 
material.

%\section{Conclusion}
To summarize, a simple solid-solid phase transformation model that predicts a morphological instability of the
interface under uniform compressional stress has been developed and investigated. The instability
is triggered by a finite jump in the elastic properties across the
interface and a concomitant jump in the free energy density. We also showed that the characteristic time of
roughening depends on the external applied stress and
the elastic parameters jump, in such a way that a higher external
compression load or a larger difference between the elastic properties of the phases shortens
the time required to roughen the interface. This result allows the roughening time and formation rate of stylolites to be estimated as a function of burial depth in sedimentary basins.

\begin{acknowledgments}
This project was funded by \textsl{Physics of Geological Processes},
a Center of Excellence at the University of Oslo. The authors are grateful to Paul Meakin for fruitful discussions and comments.
\end{acknowledgments}

%% ------------------------------------------------------------------------ %%
%
%  REFERENCE LIST AND TEXT CITATIONS
%
%% ------------------------------------------------------------------------ %%

%\end{article}

\end{document}